\documentclass{aa} 

\usepackage{natbib}
\usepackage{graphicx}
\usepackage{txfonts}
\usepackage{epsfig}
\usepackage{times}
\usepackage{rotating}
\usepackage{setspace}
\usepackage{lscape}

\def\aFe{[$\alpha/{\rm Fe}$]}

\def\Ha{${\rm H}{\alpha}$}
\def\Hb{${\rm H}{\beta}$}

\def\Mgb{{\rm Mg}\,$_b$}
\def\Fe{$\langle {\rm Fe}\rangle$}

\def\ZH{[$Z/{\rm H}$]}
\def\MgFe{[${\rm MgFe}$]$'$}

\def\oiiic{[\ion{O}{iii}]$\lambda5007$}

\def\kms{$\rm km\;s^{-1}$}

\def\spose#1{\hbox to 0pt{#1\hss}}

\def\aj{AJ}                   
\def\apj{ApJ}                 
\def\apjl{ApJ}                
\def\apjs{ApJS}               
\def\ao{Appl.~Opt.}           
\def\apss{Ap\&SS}             
\def\aap{A\&A}                
\def\aaps{A\&AS}              
\def\mnras{MNRAS}             
\def\pasp{PASP}               
 
\begin{document}

\title{Kinematic and stellar population properties of the
  counter-rotating components in the S0 galaxy NGC 1366\thanks{Based
    on observations made with ESO Telescopes at the La Silla-Paranal
    Observatory under programmes 075.B-0794 and 077.B-0767.}}

\authorrunning{L. Morelli et al.}

\titlerunning{Counter-rotating stellar components in NGC~1366}

\author{L. Morelli\inst{1,2}
        \and
        A. Pizzella\inst{1,2}
        \and
        L. Coccato\inst{3}
        \and
        E. M. Corsini\inst{1,2}
        \and
        E. Dalla Bont\`a\inst{1,2}
        \and
        L. M. Buson\inst{2}
        \and
        V. D. Ivanov\inst{3,4}
        \and\\
        I. Pagotto\inst{1}
        \and
        E. Pompei\inst{4}
        \and
        M. Rocco\inst{1}
        }

\institute{Dipartimento di  Fisica e Astronomia ``G. Galilei'', Universit\`a di Padova,
              vicolo dell'Osservatorio 3, I-35122 Padova, Italy\\
              \email{lorenzo.morelli@unipd.it}
           \and
           INAF-Osservatorio Astronomico di Padova, vicolo dell'Osservatorio~2, 
           I-35122 Padova, Italy
           \and 
           European Southern Observatory, Karl-Schwarzschild-Strasse 2, 
           D-85748 Garching bei M\"unchen, Germany
           \and
           European Southern Observatory, Avenida Alonso de C\'ordova 3107, 
           Vitacura, Casilla 19001, Santiago de Chile, Chile}

\date{\today}

\abstract{Many disk galaxies host two extended stellar components
  that rotate in opposite directions. The analysis of the stellar
  populations of the counter-rotating components provides constraints
  on the environmental and internal processes that drive their
  formation.}
{The S0 NGC 1366 in the Fornax cluster is known to host a stellar
  component that is kinematically decoupled from the main body of the galaxy. Here we successfully
  separated the two counter-rotating stellar components to
  independently measure the kinematics and properties of their stellar
  populations.}
{We performed a spectroscopic decomposition of the spectrum obtained
  along the galaxy major axis and separated the relative contribution
  of the two counter-rotating stellar components and of the
  ionized-gas component. We measured the line-strength indices of the
  two counter-rotating stellar components and modeled each of them
  with single stellar population models that account for the
  $\alpha/$Fe overabundance.}
{We found that the counter-rotating stellar component is younger, has  nearly the same metallicity, and is less $\alpha/$Fe enhanced than
  the corotating component. Unlike most of the counter-rotating
  galaxies, the ionized gas detected in NGC~1366 is neither associated
  with the counter-rotating stellar component  nor with the main
    galaxy body. On the contrary, it has a disordered distribution
  and a disturbed kinematics with multiple velocity components
  observed along the minor axis of the galaxy.}
{The different properties of the counter-rotating stellar components
  and the kinematic peculiarities of the ionized gas suggest that
  NGC~1366 is at an intermediate stage of the acquisition process,  building the counter-rotating components with some gas clouds still
  falling onto the galaxy.}


\maketitle

\section{Introduction}
\label{sec:introduction}

The photometric and kinematic analysis of nearby objects reveals that
disk galaxies may host decoupled structures on various scales, from
a few tens of pc \citep[e.g.,][]{Pizzella2002, Corsini2003, Erwin2004}
to several kpc \citep[e.g.,][]{Rubin1994, Kuijken2001, Combes2006}.
In particular, observational evidence for two stellar disks, two
gaseous disks, or for a gaseous disk and a stellar disk rotating in
opposite directions have been found on large scales in galaxies of
different morphological types \citep{Galletta1996,
  Corsini2014}. Counter-rotating stellar and/or gaseous disks occur in
$\sim30\%$ of S0 galaxies \citep{Pizzella2004, Davis2011} and in
$\sim10\%$ of spirals \citep{Kannappan2001, Pizzella2004,Corsini2012}.

Different processes have been proposed to explain the formation of a
galaxy with two counter-rotating stellar disks, and each formation
scenario is expected to leave a noticeable signature in the stellar
population properties of the counter-rotating components.
A counter-rotating stellar disk can be built from gas accreted with an
opposite angular momentum with respect to the pre-existing galaxy from
the environment or from a companion galaxy. The counter-rotating gas
settles on the galaxy disk and forms the counter-rotating stars. In
this case, the gas is kinematically associated with the counter-rotating
stellar component, which is younger and less massive than
the main body of the galaxy \citep{Thakar1996,
  Thakar1998,Algorry2014}.
Another viable, but less probable, formation process is related to the
major merger between two disk galaxies with opposite rotation. The
difference in age of the two counter-rotating components depends on
the stellar population of the progenitors and on the timescale of the
star formation triggered by the binary merger. Moreover, the two
stellar disks are expected to have a different thickness
\citep{Puerari2001, Crocker2009, Bettoni2014}.
Finally, the dissolution of a bar or triaxial stellar halo can build
two counter-rotating stellar components with similar age and mass
without involving gas. One of them is rotating in the same direction
as the bulge and disk of the pre-existing galaxy \citep[][but see also
  \citealt{Sellwood1994, Khoperskov2016}]{Evans1994}.

These predictions are difficult to be tested, since outside our Galaxy
it is a hard task to separate the single components of a composite stellar population. However, this is possible in a few galaxies because of
the difference in velocity of their extended counter-rotating stellar
components. Counter-rotating galaxies are therefore ideal laboratories
for studying how galaxies grow by episodic or continuous accretion of
gas and stars through acquisition and merging events.
\citet{Coccato2011} presented a spectroscopic decomposition
technique that allows separating the relative contribution of two
stellar components from the observed galaxy spectrum. This allows us to
study the kinematics and spectroscopic properties of individual
components independently, minimizing their cross-contamination along
the line of sight. We applied this technique to many of the galaxies
known to host counter-rotating stellar disks with the aim of
constraining their formation process \citep{Coccato2011, Coccato2013,
  Coccato2015, Pizzella2014}. In most of these cases, the available
evidence supports the hypothesis that stellar counter-rotation is the end product of
a retrograde acquisition of external gas and subsequent star
formation. Other teams developed their own algorithms for
separating the kinematics and stellar populations of
counter-rotating galaxies and found results similar to ours
\citep{Johnston2013, Katkov2011, Katkov2013, Katkov2016, Mitzkus2016}.

\begin{figure}
\includegraphics[angle=0.0,width=0.5\textwidth]{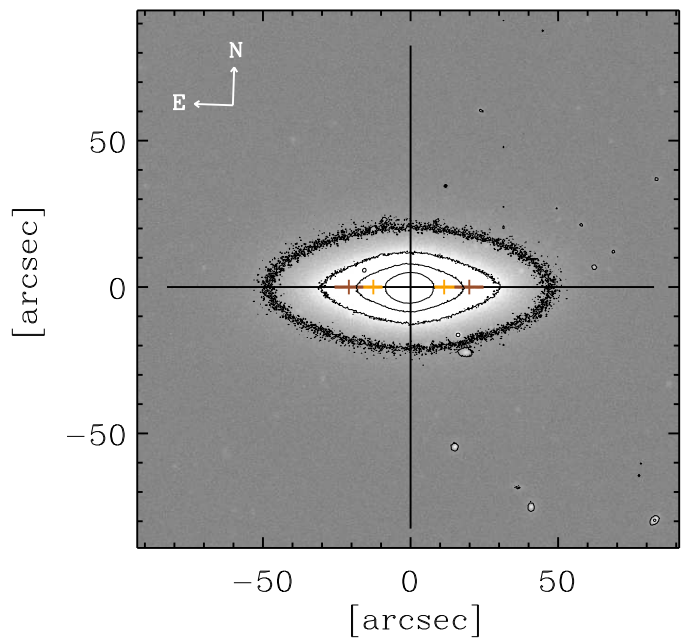}
\caption{ Contour plots in arbitrary scale of the $R$-band image of
  NGC~1366 taken from \citet{Morelli2008}. The solid lines mark the
  position of the slit along the major and minor axis of the
  galaxy. The brown and orange segments correspond to the radial bins
  where we were able to separate the two counter-rotating stellar
  components. Orientation of the field of view is given in the figure,
  and the scale is 82 pc per arcsec.}
\label{fig:image}
\end{figure}

NGC~1366 is a bright and spindle galaxy (Fig.~\ref{fig:image}) in the
Fornax cluster at a distance of 17 Mpc \citep{Ferguson1989}. It is
classified as S0$^0$ by \citet{RC3} and S0$_1$(7)/E7 by \citet{CAG}
because it has a highly inclined thin disk.  Although NGC~1366
belongs to the LGG~96 group, \citet{Garcia1993}, it does not have any
nearby bright companion and shows an undisturbed morphology. It has an
absolute total $B$ magnitude $M_{B_T}^0=-18.30$ mag, as derived from
$B_T=11.97$ mag \citep{RC3} by correcting for the inclination and
extinction given by HyperLeda \citep{Makarov2014}.  The apparent
isophotal diameters measured at a surface brightness level of $\mu_B =
25$ mag~arcsec$^{-2}$ are $2.1\times0.9$ arcmin corresponding to
$10.4\times4.5$ kpc. Its surface-brightness distribution is well
fit by a S\'ersic bulge and an exponential disk with a
bulge-to-total luminosity ratio $B/T=0.2,$ as found by
\citet{Morelli2008}. These authors detected a
kinematically decoupled stellar component that is younger than the
host bulge and has probably formed by enriched material acquired through
interaction or minor merging.
%

In this paper we revisit the case of NGC 1366 by successfully
separating the two counter-rotating components and properly
measuring the properties of their stellar populations
(Sect. \ref{sec:spectroscopy}). The analysis of the kinematics of the
stars and ionized gas and of the stellar populations is consistent
with the formation of the counter-rotating component from
external gas that is still accreting onto the galaxy
(Sect. \ref{sec:discussion}).

\section{Long-slit spectroscopy}
\label{sec:spectroscopy}

\subsection{Observations and data reduction}
\label{sec:observations}

We carried out the spectroscopic observations of NGC~1366 on 2005
January 25 with the 3.5 m New Technology Telescope (NTT) at the
European Southern Observatory (ESO) in La Silla (Chile).  We obtained
$2\times45$-minutes spectra along the major (P.A.$=2^{\circ}$) and
minor (P.A.$=92^{\circ}$) axis of the galaxy with the ESO Multi-Mode
Instrument (EMMI). It mounted a 1200 $\rm grooves\,mm^{-1}$ grating
with a 1.0 arcsec $\times$ 5.5 arcmin slit, giving an instrumental
resolution $\sigma_{\rm inst}=25$ \kms. The detector was a mosaic of
the No.~62 and No.~63 MIT/LL CCDs. Each CCD has $2048\,\times\,4096$
pixels of $15\,\times\,15$ $\rm \mu m^2$.  We adopted a $2\times2$
pixel binning. The wavelength range between about 4800 \AA\ and 5400
\AA\ was covered with a reciprocal dispersion of 0.40 
\AA\ $pixel^{-1}$ after $2\times2$ pixel binning. All the spectra were
bias subtracted, flat-field corrected, cleaned of cosmic rays, and
wavelength calibrated using standard IRAF\footnote{Image Reduction and
  Analysis Facility (IRAF) is distributed by the National Optical
  Astronomy Observatory (NOAO), which is operated by the Association
  of Universities for Research in Astronomy (AURA), Inc. under
  cooperative agreement with the National Science Foundation.}
routines. The spectra obtained along the same axis were coadded using
the center of the stellar continuum as reference. Further details
about the instrumental setup and spectra acquisition are given in
\citet{Morelli2008}. We followed the prescriptions of
\citet{Morelli2016} for the data reduction.

\subsection{Stellar and ionized-gas kinematics}
\label{sec:kinematics}

We derived the stellar kinematics along both the major and minor axis
of NGC~1366 with a single-component and with a two-components analysis
as done in \citet{Pizzella2014}.

\begin{figure*}
\includegraphics[angle=90.0,width=0.98\textwidth]{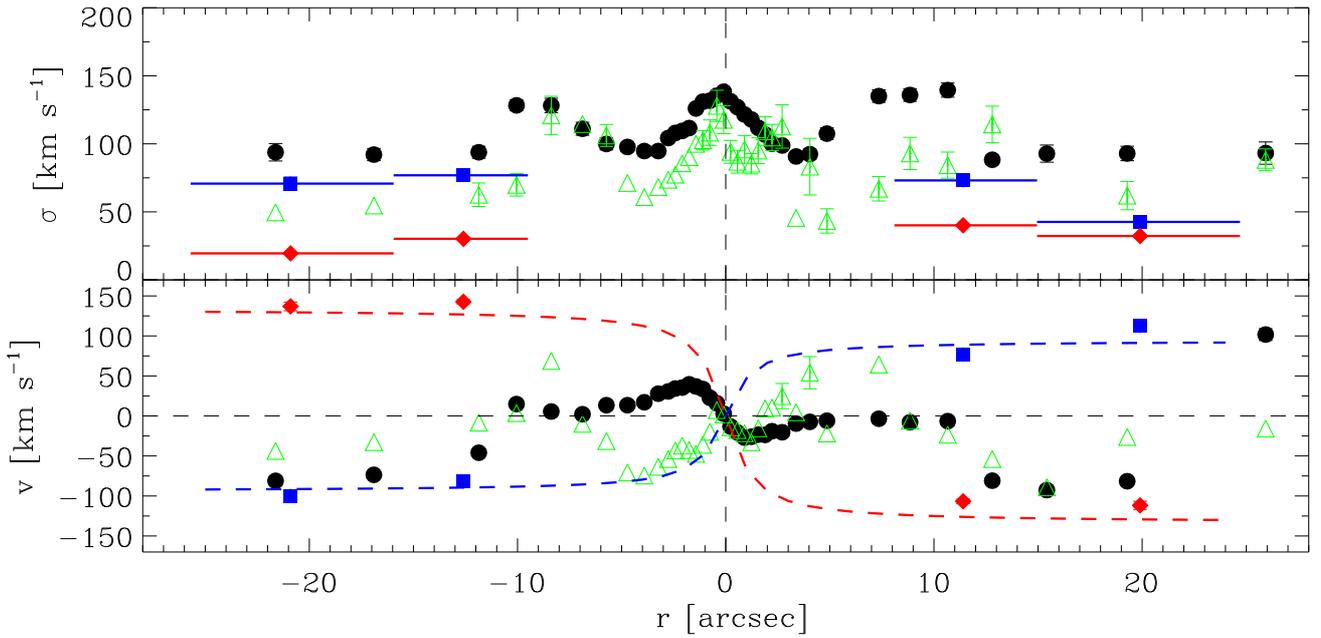}
\caption{Line-of-sight velocity dispersion (top panel) and velocity
  (bottom panel) radial profiles measured along the major axis of
  NGC~1366 for the total (black filled circles), counter-rotating
  (blue filled square), and co-rotating (red filled diamonds) stellar
  components and for the ionized gas component (green open
  triangles). Error bars smaller than symbols are not shown. The blue
  and red horizontal lines in the top panel mark the radial bins we
  adopted for measuring the counter-rotating and corotating
  components, respectively.  The blue and red dashed lines in the
  bottom panel} are a tentative indication of the velocity rotation
curves for the counter-rotating and corotating component,
respectively.
\label{fig:kin_mj}
\end{figure*}

\begin{figure*}[t!]
\includegraphics[angle=90.0,width=0.98\textwidth]{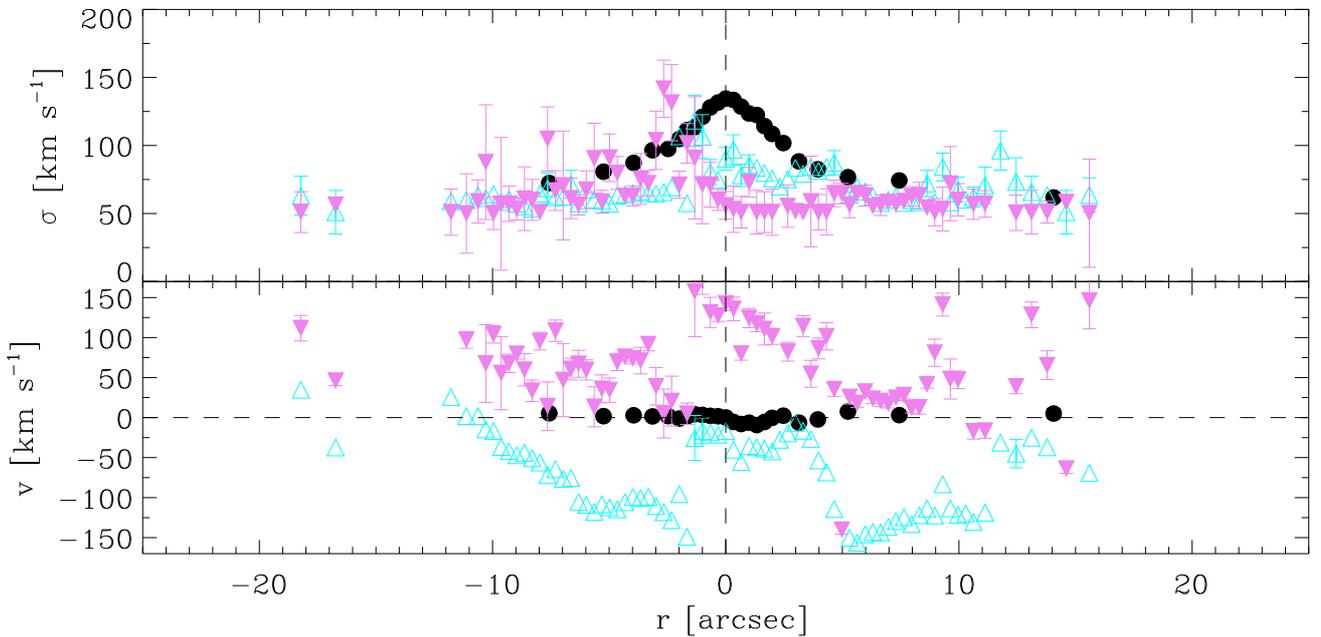}
\caption{Line-of-sight velocity dispersion (top panel) and velocity
  (bottom panel) radial profiles measured along the minor axis of
  NGC~1366 for the total stellar (black filled circles) and two
  ionized-gas components (cyan open triangles and violet filled
  triangles). Error bars smaller than symbols are not
  shown.}
\label{fig:kin_mn}
\end{figure*}

We first measured the spectra without separating the two
counter-rotating components \citep{Morelli2015}.  We used the penalized pixel fitting \citep[pPXF,][]{Cappellari2004} and gas and
absorption line fitting \citep[GANDALF,][]{Sarzi2006}
IDL\footnote{Interactive Data Language (IDL) is distributed by ITT
  Visual Information Solutions.}   codes with the ELODIE library of
stellar spectra from \citet{Prugniel2001} and adopting a Gaussian
line-of-sight velocity distribution (LOSVD) to obtain the velocity
curve and velocity dispersion radial profile along the observed
axes. We subtracted the measured velocities from the systemic velocity,
but we did not apply any correction for the slit orientation and
galaxy inclination, while we corrected the measured velocity
dispersion for the instrumental velocity dispersion.

We found a peculiar stellar kinematics along the major axis of
NGC~1366 (Fig.~\ref{fig:kin_mj}). The velocity curve is symmetric
around the center for the innermost $|r|\leq11\arcsec$. It is
characterized by a steep rise reaching a maximum of $|v|\simeq50$
\kms\ at $|r|\simeq2\arcsec$ and decreasing farther out to
$|v|\simeq0$ \kms\ at $6\la|r|\la11\arcsec$.  For $|r|\geq11\arcsec$
the spectral absorption lines clearly display a double peak that
is due to the
difference in velocity of the two counter-rotating components. The absorption lines of the two stellar populations
are so well separated that the pPXF-GANDALF procedure fit
only one of the two components. This is the reason for the shift in velocities and the drop in velocity dispersion to lower values
that we measured on both sides of the galaxy at $|r|\geq11\arcsec$
(Fig. \ref{fig:kin_mj}).  The velocities measured at large negative
and positive radii are related to the counter-rotating and corotating
component, respectively.
The velocity dispersion shows a central maximum $\sigma\simeq150$
\kms\ and decreases outwards. It rises again to peak at
$\sigma\simeq140$ \kms\ at $|r|\simeq9\arcsec$ and decreases
to a value
of $\sigma\simeq100$ \kms\ at $|r|\simeq25\arcsec$.
The combination of zero velocity with two off-centered and symmetric
peaks in the velocity dispersion of the stellar component measured
along the galaxy major axis is indicative of two
counter-rotating components. This feature shows up in the kinematics
obtained from long-slit \citep{Bertola1996,Vergani2007} and integral-field
spectroscopy \citep{Krajnovic2011,Katkov2013} when the two counter-rotating
components have almost the same luminosity and their difference in
velocity is not resolved.

We found no kinematic signature of stellar decoupling along the minor
axis of NGC~1366 (Fig.~\ref{fig:kin_mn}). The velocity curve is
characterized by $|v|\simeq0$ \kms\ at all radii, indicating that the
photometric and kinematic minor axes of the galaxy coincide with
each
other. The velocity dispersion profile is radially symmetric and
smoothly declines from $\sigma\simeq150$ \kms\/ in the center to
$\simeq60$ \kms\ at the last measured radius ($r\simeq14\arcsec$).

\begin{figure}[t!]
\centering
\includegraphics[angle=90.0,width=0.49\textwidth]{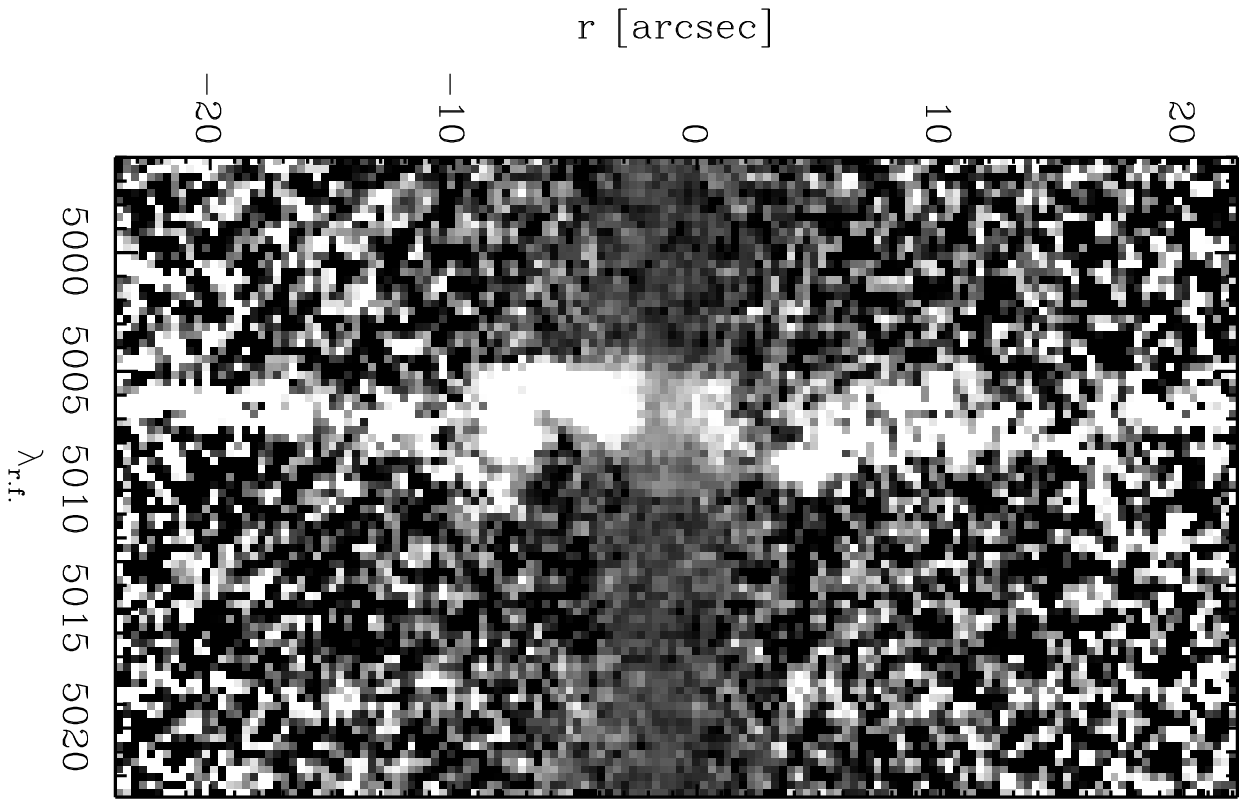}\\ 
\vspace{-0.8cm}
\includegraphics[angle=90,width=0.56\textwidth]{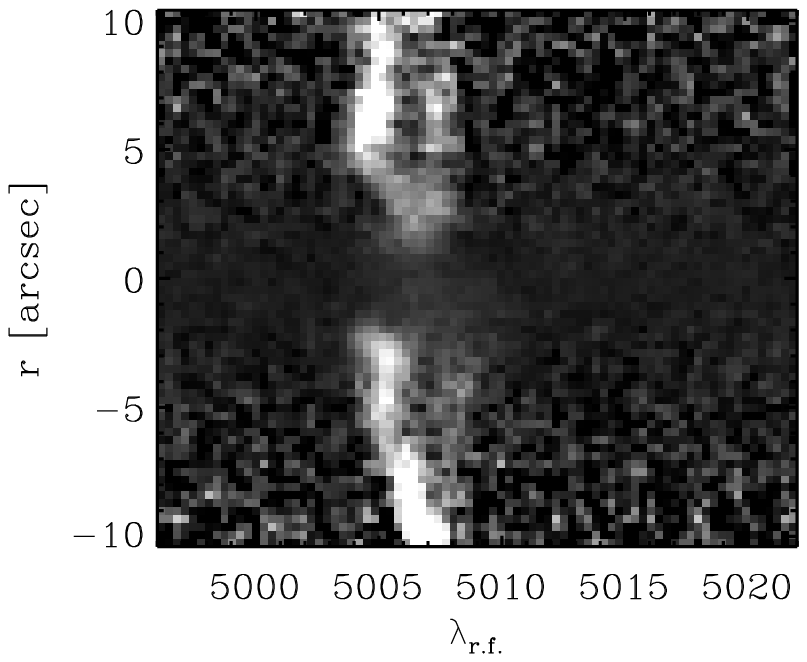}
\caption{Portion of major (top panel) and minor-axis (bottom panel)
  rest-frame spectra of NGC~1366 showing the \oiiic\/ emission line
  after subtracting the best-fitting stellar template.}
  \label{fig:oiii}
\end{figure}

Finally, we derived the kinematics of the two counter-rotating
components along the major axis at the radii where their difference in
velocity was resolved, giving rise to double-peaked absorption
lines. To reach the signal-to-noise ratio  $(S/N)$ needed to successfully perform the spectral
decomposition, we averaged the galaxy spectrum along the spatial
direction in the regions with the highest contribution of the
counter-rotating component.  We obtained a minimum $S/N \geq 30$
  per resolution element, which increases to a maximum value $S/N
  \simeq 50$ in the very central region.

\begin{figure*}[t!]
\includegraphics[angle=0.0,width=0.98\textwidth]{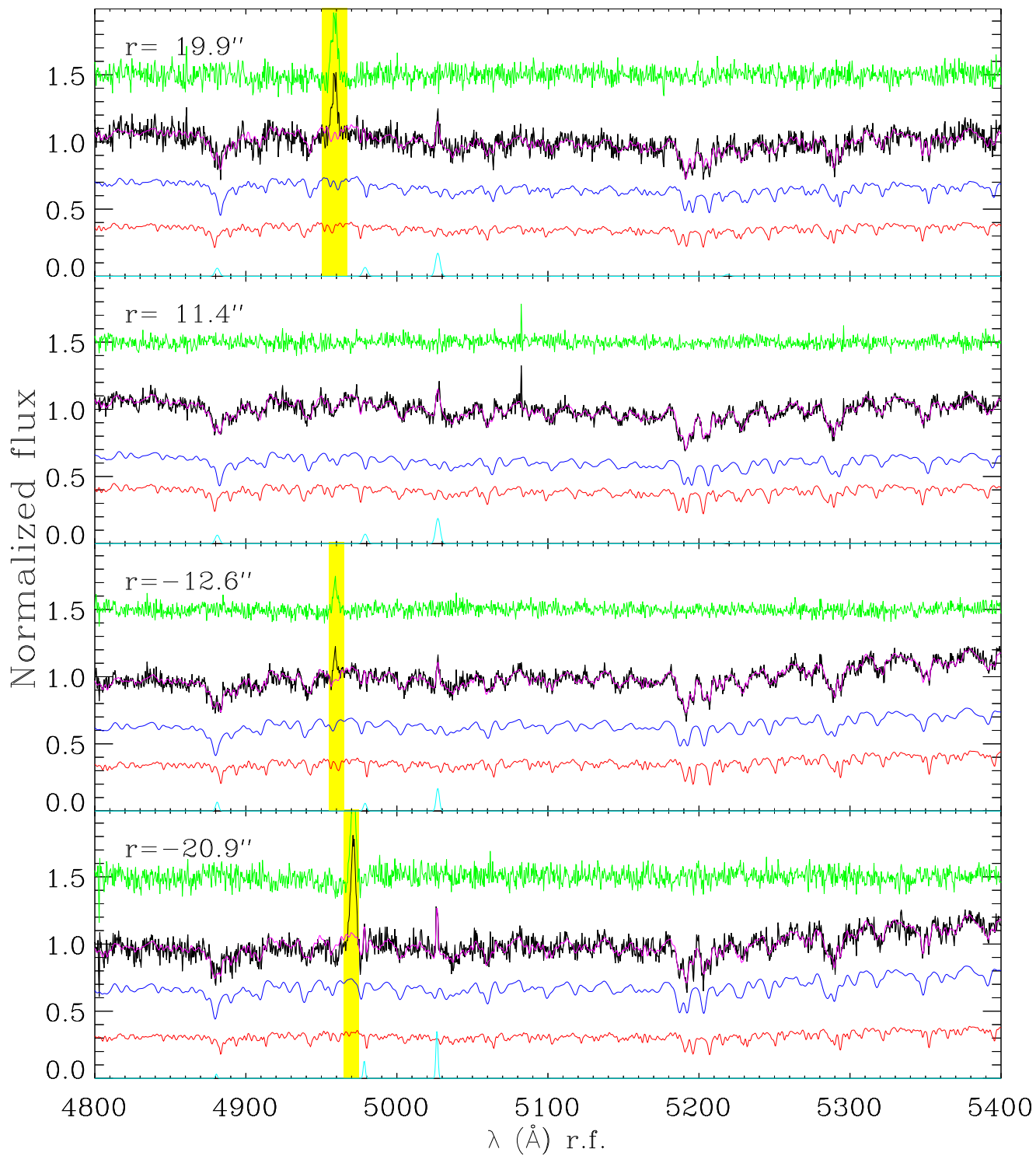}
\caption{Decomposition of the major-axis spectrum of NGC~1366 (black
  line) in the analyzed spatial bins ($r=-20.9\arcsec, -12.6\arcsec,
  11.4\arcsec$ and $19.9\arcsec$). The best-fitting model (magenta
  line) is the sum of the spectra of the corotating (red line) and
  counter-rotating stellar component (blue line) and of the
  ionized-gas component (cyan line). The normalized flux of the fit
  residual (green line) has a false zero-point for viewing
  convenience. The yellow shaded area indicates a spectral region
    masked in the fit that is due to the imperfect subtraction of the spurious
    signal, which is the result of a reflection on the EMMI CCD.}
\label{fig:2c_fit}
\end{figure*}

We performed the spectroscopic decomposition using the implementation
of the pPXF developed by \citet{Coccato2011}.  We built for each
stellar component a best-fitting synthetic template as linear
combination of the ELODIE stellar spectra. The two templates depend on
the corresponding stellar populations of the corotating and
counter-rotating components and were convolved with a Gaussian LOSVD
according to their kinematics. We added multiplicative polynomials to
deal with differences in the continuum shape of the galaxy and stellar
spectra due to flux calibration and flat fielding residuals. We also
included a few Gaussian functions to account for the ionized-gas
emission lines and generated a synthetic galaxy spectrum that
matches the
observed spectrum. The spectroscopic decomposition returns the luminosity
fraction, the line-of-sight velocity, and velocity dispersion of the
two stellar components, the line-of-sight velocity and velocity
dispersion of the ionized gas, and the two best-fitting synthetic
stellar templates to be used for the analysis of the stellar
population properties. We quantified the errors on the luminosity
  fraction, line-of-sight velocity, and velocity dispersion of the
  two counter-rotating stellar components with a series of Monte Carlo
  simulations on a set of artificial galaxy spectra, as done in \cite
  {Coccato2011}.

The decomposition of the galaxy spectrum in the radial bins at
$r=-20.9\arcsec, -12.6\arcsec, 11.4\arcsec$ and $19.9\arcsec$ are
shown in Fig.~\ref{fig:2c_fit}, and the resulting kinematics of the
corotating and counter-rotating stellar components are plotted in
Fig.~\ref{fig:kin_mj}. Corotating stars are characterized by a higher
rotation velocity ($|v|\simeq120$) \kms\ and a lower velocity
dispersion ($\sigma\simeq30$) than the counter-rotating stars that
rotate with a $|v|\simeq90$ \kms\ and have a ($\sigma\simeq80$ \kms).
The corotating and counter-rotating components contribute
$(45\pm15)\%$ and $(55\pm15)\%$ of the stellar luminosity at all the
measured radii. We converted the luminosity fraction of each component
into mass fraction using the measured ages and metallicities and
adopting the models by \citet{Maraston2005}. We derived stellar
mass-to-light ratios of $M/L=3.02$ and $M/L=1.63$ for the corotating
and counter-rotating components, respectively. From these quantities
we found that the stellar mass fractions of the corotating and
counter-rotating components are $60\%$ and $40\%$, respectively.

A comparison between the stellar and ionized-gas velocity curves
indicates that the gas is disturbed and is not associated with one of
the two counter-rotating components. In fact, the gas rotates in the
same direction and with a velocity amplitude close to that of the
stellar component at small ($|r|\la1\arcsec$) and large radii
($|r|\geq11\arcsec$). A broad feature is clearly visible in the
  gas structure at $|r|\simeq7-10\arcsec$ along the major axis
  (Fig.~\ref{fig:oiii}). Although the \oiiic\/ emission line has a
  broad profile (Fig.~\ref{fig:2c_fit}), there is no clear evidence
  for a double peak. The wavelength range of our spectra does not
  cover the \Ha\/ region, which prevents us form building a complete diagnostic
  diagram to properly distinguish between the different excitation
  mechanisms of the ionized gas. However, the high value of log(\oiiic
  /\Hb)$\,\simeq\,1.5$ favors the shocks as excitation mechanism.

 We detected two ionized-gas rotating components along the galaxy
 minor axis as it results from the double-peaked \oiiic\ emission line
 shown in Fig.~\ref{fig:oiii}. We independently measured the brighter
 emission line at lower velocities and the fainter emission line at
 higher velocities. Their velocity and velocity dispersion are shown
 in Fig.~\ref{fig:kin_mn}. The two gas components have a systematic
 and almost constant offset in velocity with respect to the stellar
 component, suggesting the presence of more gas clouds along the line
 of sight.  We prefer this interpretation to the idea of having two
 gas components with mirrored asymmetric distributions with a brighter
 and a fainter side and giving rise to an X-shaped \oiiic\ emission
 line.  The gas velocity dispersion is typically $\sigma_{\rm gas} <
 100$ \kms\ and mostly $\sigma_{\rm gas} \simeq 50$ \kms\ along both
 axes after correcting for the instrumental velocity dispersion.

\subsection{Stellar populations}
\label{sec:populations}

\begin{figure}
\centering
\includegraphics[angle=90.0,width=0.49\textwidth]{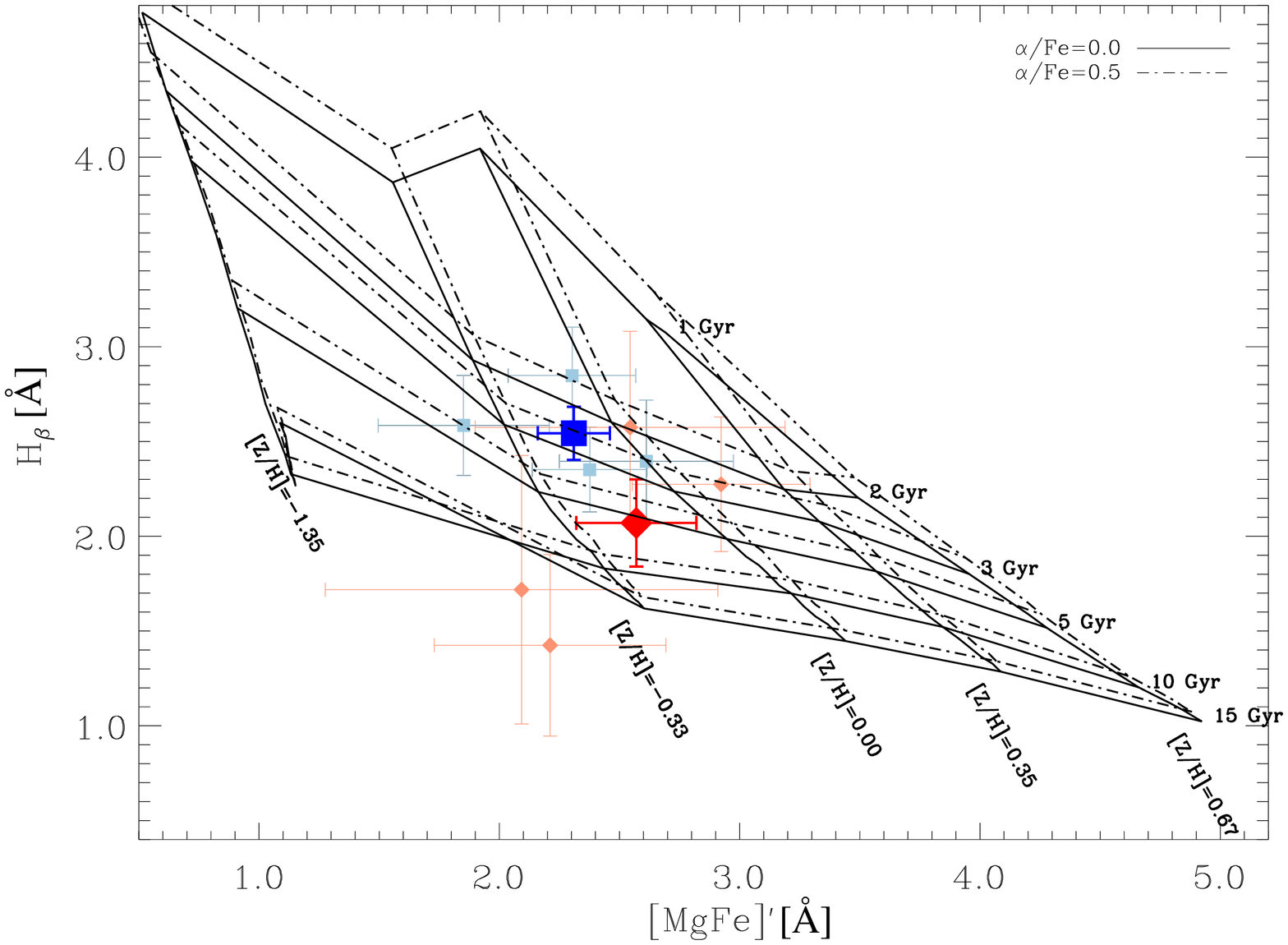} \\
\includegraphics[angle=90.0,width=0.49\textwidth]{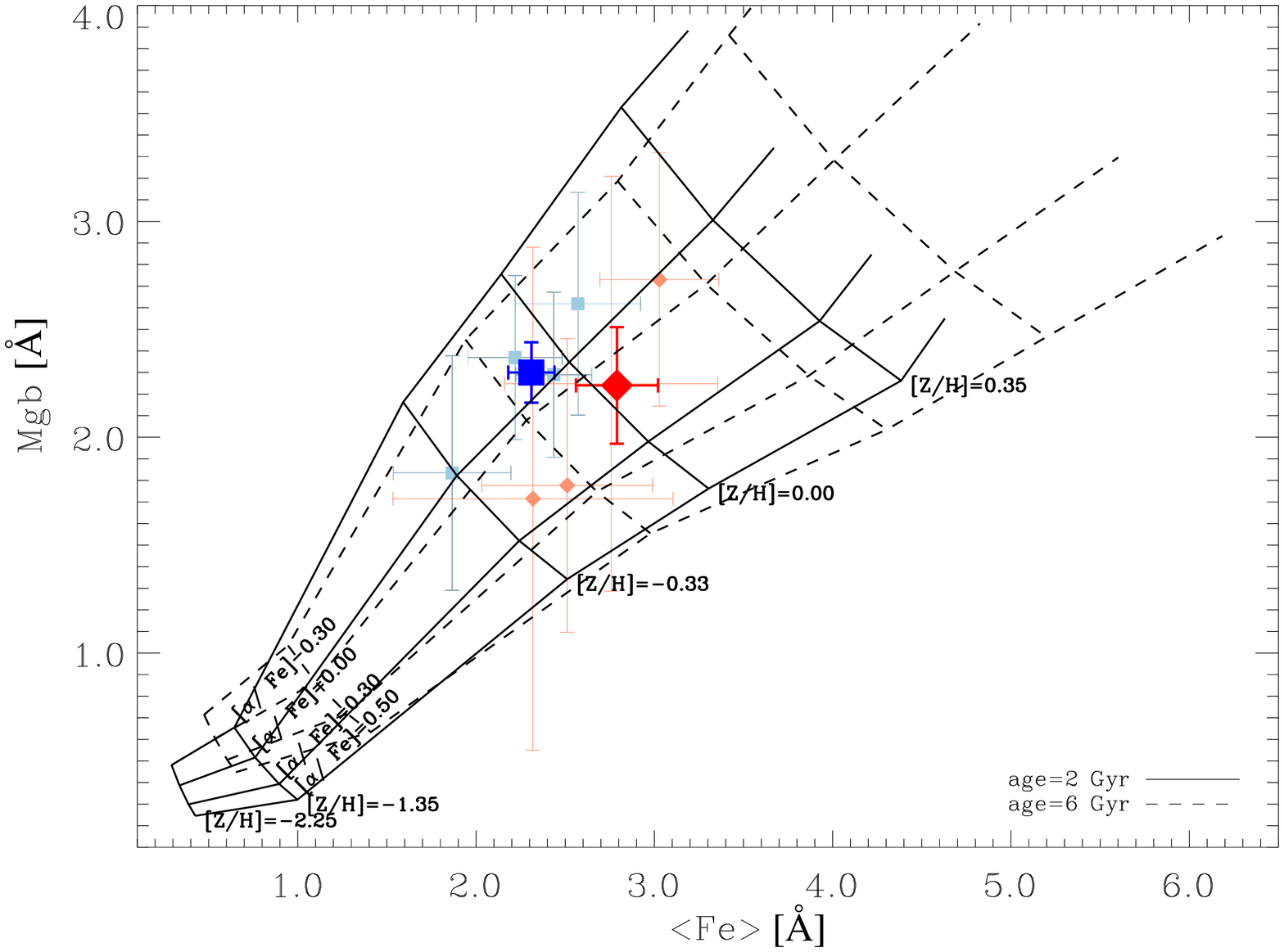}
\caption{Values of \Hb\ and \MgFe\ line-strength indices (top panel)
  and \Fe\ and \Mgb\ line-strength indices (bottom panel) for the
  corotating (small red diamonds) and counter-rotating stellar
  component (small blue squares) measured along the major axis of
  NGC~1366 ($r=-20.9\arcsec, -12.6\arcsec, 11.4\arcsec,$ and
  $19.9\arcsec$). The larger symbols correspond to the averaged
  line-strength indices for the two stellar components. The lines
  indicate the model predictions by \citet{Thomas2003} for different
  \aFe\ ratios (top panel) and ages (bottom panel).}
  \label{fig:griglie}
\end{figure}

We measured the Lick line-strength indices \citep[as defined
  in][]{Gorgas1990, Worthey1994, Thomas2003} of the corotating and
counter-rotating components on the best-fitting synthetic templates
and derived the age, metallicity, and \aFe\ ratio of the corresponding
stellar population as in \citet{Morelli2012}. We derived the
  errors on the equivalent widths of the line-strength indices of the
  two counter-rotating stellar components with a series of Monte Carlo
  simulations on a set of artificial galaxy spectra as done in \cite
  {Coccato2011}. We report the measurements in
Table~\ref{tab:lick2c} and compare them to the line-strength indices
predicted for a single stellar population that accounts for the
$\alpha/$Fe overabundance by \citet{Thomas2003} in
Fig.~\ref{fig:griglie}. We obtained the stellar population properties
of both components from the line-strength indices averaged on the two
galaxy sides. They are given in Table~\ref{tab:agemetalfa} together
with the relative luminosity of the corotating and counter-rotating
components.

\begin{table}[t!]
\caption{Line-strength indices of the corotating and counter-rotating
  stellar components of NGC~1366.}
\begin{small}
\begin{tabular}{rcccc}
\hline
\noalign{\smallskip}
\multicolumn{1}{c}{$r$} &
\multicolumn{1}{c}{\Hb} &
\multicolumn{1}{c}{\Mgb} &
\multicolumn{1}{c}{Fe$_{5270}$} &
\multicolumn{1}{c}{Fe$_{5335}$} \\
\noalign{\smallskip}
\multicolumn{1}{c}{[\arcsec]} &
\multicolumn{1}{c}{[\AA]} &
\multicolumn{1}{c}{[\AA]} &
\multicolumn{1}{c}{[\AA]} &
\multicolumn{1}{c}{[\AA]} \\
\noalign{\smallskip}
\hline
\noalign{\smallskip} 
\multicolumn{5}{c}{Corotating component} \\
\noalign{\smallskip} 
$-20.9$ & $1.72\pm0.71$  & $2.32\pm0.79$ & $2.11\pm0.89$ & $1.32\pm0.80$\\
$-12.6$ & $1.43\pm0.48$  & $2.51\pm0.48$ & $2.16\pm0.47$ & $1.39\pm0.49$\\
$ 11.4$ & $2.27\pm0.36$  & $3.03\pm0.33$ & $2.94\pm0.39$ & $2.52\pm0.44$\\
$ 19.9$ & $2.57\pm0.51$  & $2.76\pm0.60$ & $2.47\pm0.68$ & $2.02\pm0.68$\\
\noalign{\smallskip} 
\multicolumn{5}{c}{Counter-rotating component} \\
\noalign{\smallskip} 
$-20.9$ & $2.40\pm0.32$ & $2.57\pm0.32$ & $2.70\pm0.38$ & $2.54\pm0.35$\\ 
$-12.6$ & $2.85\pm0.26$ & $2.22\pm0.26$ & $2.42\pm0.27$ & $2.32\pm0.27$\\
$ 11.4$ & $2.35\pm0.22$ & $2.44\pm0.21$ & $2.35\pm0.25$ & $2.23\pm0.29$\\ 
$ 19.9$ & $2.59\pm0.26$ & $1.87\pm0.33$ & $1.84\pm0.39$ & $1.83\pm0.38$\\ 
\noalign{\smallskip}
\hline
\end{tabular}
\end{small}
\label{tab:lick2c}
\end{table}

\begin{table}
\caption{Properties of the stellar populations of the corotating and
  counter-rotating stellar components of NGC~1366.}
\begin{center}
\begin{small}
\begin{tabular}{lcrrr}
\hline
\noalign{\smallskip}
\multicolumn{1}{c}{Component} &
\multicolumn{1}{c}{$L/L_{\rm T}$} &
\multicolumn{1}{c}{Age} &
\multicolumn{1}{c}{\ZH} &
\multicolumn{1}{c}{\aFe} \\
\noalign{\smallskip}
\multicolumn{1}{c}{} &
\multicolumn{1}{c}{} &
\multicolumn{1}{c}{[Gyr]} &
\multicolumn{1}{c}{[dex]} &
\multicolumn{1}{c}{[dex]} \\
\noalign{\smallskip}
\hline
\noalign{\smallskip}
Corotating      &$0.45$ &$ 5.6\pm2.7$ & $ -0.18\pm0.16$ & $ 0.08\pm0.13$ \\
Counter-rotating &$0.55$ &$ 2.6\pm0.5$ & $ -0.16\pm0.11$ & $-0.07\pm0.08$ \\
\noalign{\smallskip}
\hline
\noalign{\medskip}
\end{tabular}
\end{small}
\label{tab:agemetalfa}
\end{center}
\end{table}

The comparison of the averaged age values suggests that the
counter-rotating component is significantly younger (Age$\,=\,2.6$
Gyr) than the corotating component (age$\,=\,5.6$ Gyr). The two
averaged
metallicities are both subsolar and similar to each other (\ZH$\,=-0.16$ and
$-0.18$ dex for the counter-rotating and corotating components,
respectively). However, the large scatter in the metallicity
measurements of the corotating component does not allow us to give a
firm conclusion. At face value, the subsolar \aFe\ ratio of the
counter-rotating component (\aFe$\,=\,-0.07$ dex) points to a longer
star-formation timescale than that of the corotating component,
which is characterized by a supersolar \aFe\ ratio (\aFe$\,=\,0.08$
dex).

\section{Discussion and conclusions}
\label{sec:discussion}

There is no morphological or photometric evidence that NGC~1366 is
hosting two counter-rotating stellar components. NGC~1366 is
characterized by an undisturbed morphology with no sign of recent
interaction with small satellites or companion galaxies of similar
size \citep{Morelli2008}. This is common for most
of the counter-rotating galaxies since their environment does not
appear statistically different from that of normal galaxies,
see \citet{Bettoni2001}. In addition, the surface brightness distribution
of NGC~1366 is remarkably well fitted by a S\'ersic bulge and an
exponential disk with no break at any radius \citep{Morelli2008}.

We provided the spectroscopic evidence of two
counter-rotating stellar components with a high rotation velocity and
low velocity dispersion ($v/\sigma\simeq2$) that give almost the same
contribution to the galaxy luminosity. We infer that they have a similar
scale length from the constant slope of the exponential
surface-brightness radial profile outside the bulge-dominated region
as in NGC~4138 \citep{Jore1996, Pizzella2014} and NGC~4550
\citep{Rix1992, Coccato2013, Johnston2013}. These kinematic and
photometric properties support the disk nature of the two components.

The stellar population of the corotating component is characterized
by an older age, consistent with that of bulge \citep[$5.1\pm1.7$
  Gyr,][]{Morelli2008}, subsolar metallicity, and almost solar
$\alpha$/Fe enhancement. This suggests a formation timescale of a few
Gyr that occurred at the time of the galaxy assembly. The
counter-rotating stellar component is remarkably younger with lower
$\alpha$/Fe enhancement and subsolar metallicity. The metallicity and
age values obtained for the two components are consistent within the
errors with the results obtained by \citet{Morelli2008} on the galaxy
integrated light when considering its strong radial gradients of
stellar population properties. Therefore, the counter-rotating stellar
component could be the end result of a slower star formation process
that occurred in a disk of gas accreted by a preexisting galaxy and
settled onto retrograde orbits. However, unlike most of previously
studied cases \citep[e.g.,][]{Johnston2013,Pizzella2014,Coccato2015,Katkov2016},
the ionized gas of NGC~1366 is not associated with the
counter-rotating stellar component. It has peculiar kinematics with multiple velocity components
along the minor axis with different gas clouds along
the line of sight. The kinematic mismatch between the ionized
gas and counter-rotating stellar component complicates the scenario of
gas accretion followed by star formation.

The most obvious possibility is to consider an episodic gas
accretion. The first event of capture of external gas occurred $\sim3$
Gyr ago and built the counter-rotating stellar component. It was
followed by a subsequent event that is still ongoing at
present. However, this rises the question about the origin of the
newly supplied and kinematically decoupled gas since there is no clear
donor candidate in the neighborhood of NGC~1366. This leaves us with
the possibility of the acquisition of small gas clouds coming either
from the environment or from the internal reservoir inside the galaxy
itself.  When external gas is captured in distinct clouds, it settles
onto the galaxy disk in a relatively short time \citep[$\sim1$
  Gyr,][]{Thakar1997, Algorry2014,Mapelli2015}. In this case, NGC~1366
could be an object caught at an intermediate stage of the acquisition
process, before its configuration becomes stable. It is interesting to
note that this could also have occurred in galaxies with gas associated
with the counter-rotating stellar component. Without clear evidence of
ongoing star formation or very young stars, the counter-rotating
stellar component could be the result of a past acquisition of gas
coming from the same reservoir that provides the counter-rotating gas we
observe at present.

An intriguing alternative was explored by \citet{Crocker2009}.  They
showed the time evolution of the distribution and kinematics of gas and
stars in a set of numerical simulations aimed at investigating the
formation of the stellar counter-rotating disks of NGC 4550 from a
binary merger.  One Gyr after the merger, while the stars have
settled in two counter-rotating disks with a relatively regular
kinematics, the gas distribution still remains rather disordered with
a disturbed kinematics. However, this configuration is not stable, and
the gas tends to a more regular configuration between 1 and 2 Gyr from
the merging event. The structure and stellar populations properties of
the counter-rotating components of NGC~1366 are somewhat different from
those of NGC~4550 for a direct comparison of our results with the
simulations by \citet{Crocker2009}, and dedicated simulations are
needed for a firmer interpretation of this galaxy in terms of a binary
merger.

These speculations need further evidence since the available
spectroscopic data are not conclusive. To date, NGC~1366 is a unique
example, and it may become a corner stone for understanding the formation
of counter-rotation in relatively isolated and undisturbed
galaxies. Mapping the ionized-gas distribution and kinematics of
NGC~1366 with integral-field spectroscopy is a crucial complement for
the present dataset and is necessary to distinguish between different
scenarios and address the question of the origin of the gas. In the
case of a episodic gas acquisition, we expect to see a clear
morphological and kinematic signature of the incoming gas without a
counter-part in the stellar distribution. In contrast, in the case
of a galaxy binary merger, we expect to observe a morphological
association between the distribution of stars and gas, a regular
velocity field for the two counter-rotating stellar disks, and an
irregular velocity field for the ionized gas.

\begin{acknowledgements}

We benefited from discussion with Roberto P. Saglia. This work was
supported by Padua University through grants 60A02-5857/13,
60A02-5833/14, 60A02-4434/15, and CPDA133894. LM and EMC acknowledge
financial support from Padua University grants CPS0204 and
BIRD164402/16, respectively. LM is grateful to the ESO Scientific
Visitor Programme for the hospitality at ESO Headquarters while this
paper was in progress.  This research made use of the HyperLeda
Database (http://leda.univ-lyon1.fr/) and NASA/IPAC Extragalactic
Database (NED) which is operated by the Jet Propulsion Laboratory,
California Institute of Technology, under contract with the National
Aeronautics and Space Administration (http://ned.ipac.caltech.edu/).

\end{acknowledgements}

\bibliographystyle{aa} 

\end{document}